\documentclass[twocolumn,aps,prl,floats,amsfonts]{revtex4}
\usepackage{graphicx}

\newcommand {\be}{\begin{equation}}
\newcommand {\ee} {\end{equation}}
\newcommand {\ba}{\begin{eqnarray}}
\newcommand {\ea} {\end{eqnarray}}

\begin{document}


\title{Incompressible liquid state of rapidly-rotating bosons
at filling factor $3/2$}
\author{E.H. Rezayi$^{1}$, N. Read$^{2}$ and N.R. Cooper$^{3}$}
\address{$^1$Department of Physics, California State University,
Los Angeles, CA 90032}
\address{$^2$Department of Physics, Yale University, P.O. Box
208120, New Haven, CT 06520-8120}
\address{$^3$T.C.M. Group, Cavendish Laboratory, Madingley Road,
Cambridge, CB3 0HE, United Kingdom}


\begin{abstract} Bosons in the lowest Landau level, such as
rapidly-rotating cold trapped atoms, are investigated numerically
in the specially interesting case in which the filling factor
(ratio of particle number to vortex number) is $3/2$. When a
moderate amount of a longer-range (e.g.\ dipolar) interaction is
included, we find clear evidence that the ground state is in a
phase constructed earlier by two of us, in which excitations
possess non-Abelian statistics.
\end{abstract}
\maketitle


There is increasing interest in rapidly-rotating ultra-cold atoms
in a trap. For bosons rotating at a moderate frequency, a vortex
lattice is observed \cite{vortex-exp}. At sufficiently high
rotation frequency (close to the natural frequency of a harmonic
trap), it is expected that the lattice melts and is replaced by a
series of highly-correlated liquids that can be related to
fractional quantum Hall (QH) states. The crucial parameter is the
ratio $\nu=N/N_V$ of the number of atoms, $N$, to the number $N_V$
of quantized vortices that would pierce the cloud if it were a
Bose condensate. This ratio corresponds to the Landau level
filling factor $\nu$ in the fractional quantum Hall effect.
Previous theoretical work on this regime has emphasized the
importance of the lowest Landau level (LLL) when interactions are
weak and the temperature is low, and pointed out that in this
restricted space of states the Laughlin $\nu=1/2$ state
\cite{laugh} is the exact ground state for the standard
``contact'' form of interaction \cite{wgs}. Later, evidence of a
sequence of correlated liquids was found at $\nu=k/2$, $k=1$, $2$,
$3$, \ldots for $\nu\leq \nu_c$ \cite{cwg}, and a vortex lattice
at $\nu>\nu_c$, with $\nu_c\simeq 6$--$10$ \cite{cwg,shm}. The
correlated liquids were found \cite{cwg} to have large overlaps
with states proposed by two of us (RR) some time ago \cite{rr99}.
The $k=1$ liquid is the familiar Laughlin state, while $k=2$ is
the Moore-Read (MR) paired state \cite{mr}. More recent work
strengthens the case for the MR state at $\nu=1$, and provides
some evidence for liquid states at still other filling factors not
in the sequence $\nu=k/2$ \cite{rj}, however results at $\nu=3/2$
were inconclusive. Meanwhile, experiments are approaching the LLL
regime \cite{schweik}, though the filling factors are still $\gg
\nu_c$. Very recently, Bose condensation has been achieved in
atoms with a large magnetic dipole moment \cite{Cr}, and the
effect of dipolar interactions on the vortex lattices and on the
quantum fluids at low filling factors has been investigated
\cite{crs}.

In this paper, we study numerically the next member, $\nu=3/2$, of
the RR sequence for system sizes larger than in Refs.\
\cite{rr99,cwg,rj}. We consider the s-wave (contact) interaction
in the LLL, and the effect of adding a longer-range component such
as the dipolar interaction (for dipole moments oriented parallel
to the rotation axis). This case, $k=3$, is of interest for
several reasons: For $k>1$, the quasiparticle excitations of each
of the sequence of states introduced in Ref.\ \cite{mr,rr99} have
the fascinating property of non-Abelian statistics \cite{mr}. This
makes these states even more exotic than the Laughlin \cite{laugh}
and hierarchy/composite-fermion \cite{hier} states, in all of
which the quasiparticles have fractional, but Abelian, statistics.
$k=3$ is the smallest $k$ value in the RR sequence for which the
non-Abelian statistics support universal quantum computation
\cite{freedman}. For this filling factor, $\nu=3/2$, there is also
an alternative candidate, which is a hierarchy/composite-fermion
phase. Using both the sphere and the torus geometries (to avoid
edge effects), we find clear evidence that the boson system at
$\nu=3/2$ with a moderate amount of longer-range interaction added
is in the phase described by our trial states, and hence is
non-Abelian; this may also be the case for the pure s-wave
interaction, as suggested by Ref. \cite{cwg}. The evidence comes
from the energy spectrum on the torus, which shows a ground-state
doublet with very large overlaps with the RR trial states, and a
relatively large gap to higher excited states, and from the
two-particle correlation function.

The conventional effective interaction Hamiltonian for the atoms,
representing s-wave scattering at low momentum,
is%
\be H_{\rm s}=g\sum_{1\leq i<j\leq N} \delta^3({\bf r}_i-{\bf
r}_j),\ee %
where $g=4\pi \hbar^2 a/M$ ($a$ is the s-wave scattering length,
and $M$ is the mass of an atom). 
It is of interest to consider also electric or
magnetic dipole interactions between the atoms, of the form%
\be%
H_{\rm dip} = C_d \sum_{1\leq i<j\leq N} \frac{{\bf p}_i\cdot{\bf
p}_j -3({\bf n}_{ij}\cdot{\bf p}_i)({\bf n}_{ij}\cdot{\bf
p}_j)}{|{\bf r}_i-{\bf
r_j}|^3},\ee%
where ${\bf n}_{ij}=({\bf r}_i-{\bf r}_j)/|{\bf r}_i-{\bf r}_j|$,
and the ${\bf p}_i$s are {\em unit} vectors representing the
(fixed) dipole moments. We will assume that the dipole moments are
parallel to the $3$ axis; then any $\delta$-function term that may
accompany the dipolar interaction can be absorbed into the s-wave
interaction term in $H_{\rm int}=H_{\rm s}+H_{\rm dip}$.

We work in an axially symmetric harmonic trap, with frequencies
$\omega_3$, $\omega_\perp$ for motion respectively along, and
perpendicular to, the symmetry ($3$-) axis. For a system of bosons
rotating rapidly about the $3$-axis, we will restrict the atoms to
the lowest Landau level (LLL) states, which are the
single-particle states of lowest energy for each value $m\geq 0$
of angular momentum $L_3/\hbar$ (and thus have no excitation along
the $3$-axis). This is based on the assumption that interactions
are weak, that is $g\bar{n}$ and $C_d\bar{n}$ ($\bar{n}$ is the
typical density of particles in the drop of atoms) are small
compared with the energy for excitation to higher states, that is
with $2\hbar\omega_\perp$ and $\hbar\omega_3$ \cite{wgs,cwg}. The
LLL states can be represented by functions in two dimensions,
$u_m(z)=z^m e^{-|z|^2/4}/\sqrt{2\pi 2^m m!}$, where $z=x+iy$
(henceforth we use units in which the magnetic length $\ell_B$ and
$\hbar$ are 1 \cite{ellb}).

When working on the sphere \cite{hald83}, there is a finite number
$N_V$ of flux quanta penetrating the surface of radius
$R=\sqrt{N_V/2}$, and there are $N_V+1$ single-particle states,
which have wavefunctions $u_m(z)\propto
z^m/(1+|z|^2/4R^2)^{1+N_V/2}$, $m=0$, \ldots, $N_V$ in terms of
the coordinate $z$ in the plane (from stereographic projection of
the sphere). In the torus geometry (i.e.\ periodic boundary
conditions on a parallelogram), there are $N_V$ LLL basis states
when $N_V$ flux pierce the system.

In the subspace of LLL states in the infinite plane, the
interaction Hamiltonian $H_{\rm int}$ can be represented by the
pseudopotentials $V_m$, $m=0$, $2$, \ldots. $V_m$ is the
interaction energy for a single pair of bosons of relative angular
momentum $m$ (only even $m$ are relevant for bosons)
\cite{hald83}. The pseudopotentials can be obtained \cite{crs}
from the analytic expressions in the limit (for simplicity) of
vanishing thickness of the two-dimensional fluid,
$\ell_3/\ell_\perp\to 0$ \cite{ellb}. In this limit an infinite
constant has to be absorbed into $g$. The $m\neq0$
pseudopotentials are determined entirely by the dipolar
interaction, and hence their ratios are fixed; they are plotted as
an inset in Fig.\ \ref{fig:overlaps} below. $V_0$ can be treated
as an independent parameter, so that the ratio of $V_2/V_0$ is a
dimensionless parameter characterizing the interaction apart from
one overall energy scale. We also studied an interaction in which
$V_m$ for $m>2$ is set to zero. This description of the
interaction can be extended to the sphere (also using rotation
symmetry), and to the torus.

\begin{figure}
{\centering \includegraphics[width=3.0in]{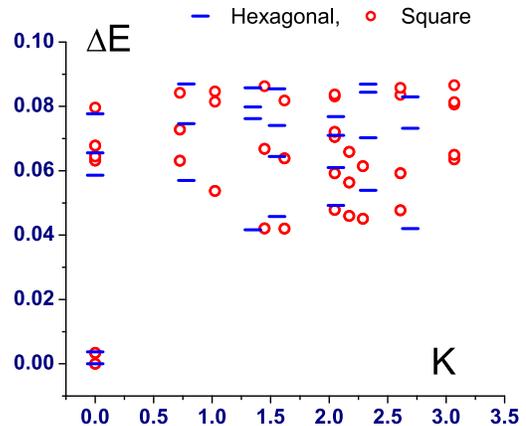} }
\caption{ \label{fig:specpbc} Low-lying spectrum for $18$ bosons
for dipolar model $H_{\rm int}$ with $V_2/V_0=0.380$ on the torus vs.\ the
pseudomomentum $K=|{\bf K}|$, for two different unit cells.}
\end{figure}

A $p$-body $\delta$-function interaction,%
\be%
H_p=\frac{W_p}{p!}\sum_{i_1,i_2,\ldots,i_p=1}^N\delta^2(z_{i_1}-z_{i_2})\cdots
\delta^2(z_{i_{p-1}}-z_{i_p}),\ee projected to the LLL is also of
interest (this form is correct for the plane and torus geometries,
and for the sphere there is a corresponding rotationally-invariant
form). The parafermion states found in ref.\ \cite{rr99} are
unique, exact zero-energy eigenstates of such interactions for
$p=k+1$ when $N$ is divisible by $k$ and $N_V=2N/k -2$ (on the
sphere), so that $\nu=\lim_{N\to\infty} N/N_V=k/2$. These states
serve as trial wavefunctions with which the exact ground states
for $H_{\rm int}$ can be compared. They represent incompressible
liquid phases, in which the excitations enjoy non-Abelian
statistics for $k>1$ (however, we will later suggest that a caveat
to this statement is required).

Before turning to our results, we review some aspects of the LLL
on a torus \cite{hald85}. On the torus at $\nu=N/N_V=3/2$,
translational symmetry implies that all energy eigenstates possess
a trivial center-of-mass degeneracy of $2$, which is exact for any
size system, and also that there is a conserved
pseudomomentum $\bf K$ \cite{hald85}, which is a vector lying in a
certain Brillouin zone. In the RR phases, the ground states have a
net degeneracy $k+1$ in the thermodynamic limit, which is
connected with the non-Abelian statistics \cite{mr,rr99}. For
$k=3$, this 4-fold degeneracy is made up of the trivial factor $2$
(which is always discarded in numerical studies), together with a
further $2$-fold degeneracy, which in general becomes exact only
in the thermodynamic limit; all these ground states have ${\bf
K=0}$. (For the $4$-body interaction, the fourfold degeneracy is
exact for any size, as the ground states have exactly zero
energy.) By contrast, the hierarchy/composite-fermion ground state
for bosons at $\nu=3/2$ possesses only the trivial $2$-fold
degeneracy. Then for an incompressible fluid on the torus, the
spectrum of a sufficiently large system in one of these two phases
should exhibit a nearly degenerate pair of ground-states (resp., a
single ground state) at $\bf K = 0$, separated by a clear gap from
a region of many states at higher energy eigenvalues.

In Fig.\ \ref{fig:specpbc} we show the spectrum for 18 particles
on the torus for $H_{\rm int}$ with $V_2/V_0=0.380$ (energies are in units of
$C_d$ for the dipolar $H$). A clear doublet can be seen at ${\bf K}=0$, and the
rest of the spectrum is well-separated compared with either the
spacing of the doublet, or the spacing among the levels above the
gap. This spectrum leaves little doubt that the fluid is
incompressible. The results are similar for the two geometries of
the torus (hexagonal and square unit cells) shown. Spectra at
nearby $V_2/V_0$, and at other aspect ratios, are similar.

\begin{figure}
{\centering \includegraphics[width=3.0in]{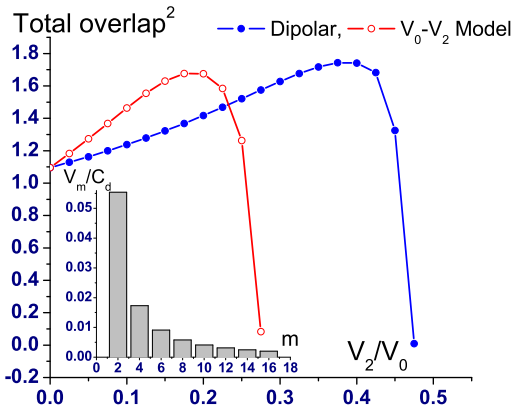} }
\caption{ \label{fig:overlaps} The total overlap-squared as a
function of $V_2/V_0$ on the torus (square unit cell), for $18$ particles, for both
the dipolar and the $V_0$, $V_2$-only model interactions. The
inset shows the $m\neq0$ pseudopotentials for the dipolar
interaction.}
\end{figure}

The overlap of each of the two low-lying $\bf K=0$ states with the
trial ground state doublet of the 4-body interaction can be
calculated. We may add the squares of these overlaps to obtain the
total overlap-squared (which is at most 2) of the two-dimensional
subspace of $H_{\rm int}$ with that of $H_4$. This total
overlap-squared is plotted versus $V_2/V_0$ for both the $V_0$
plus dipole interaction and the $V_0$, $V_2$-only interaction in
Fig.\ \ref{fig:overlaps} (the total is made up of roughly equal
contributions from each of the two low-lying states, across the
whole range of $V_2/V_0$ values). For sufficiently large
$V_2/V_0$, both cases show a very abrupt drop in overlap which is
due to a phase transition. Beyond this point, the low-lying
spectrum shows indications that the ground state in the
thermodynamic limit breaks translational symmetry. The transition
is probably first-order, but does not occur here by crossing of
energy levels of different symmetry, because a crystal and the RR
fluid both have ground states at ${\bf K}=0$. This transition is
similar to that found in other recent work at $\nu=1/2$ \cite{crs}
and will not be pursued here. We note, however, the similarity of
these overlap curves with those for $\nu=1/2$ in the case of
fermions as the interaction is changed, which show similar trends
and similar curves \cite{rh}.

The $N=18$ overlaps should be viewed as significantly large. Note
that for a random vector in $D$ dimensions, the probability of
obtaining an overlap with a given vector (or two-dimensional
subspace) greater than some given value is of order $e^{-D}$ as
$D\to\infty$. The probability of obtaining similar overlaps for
two random vectors is of order the square of this. In our case,
the $\bf K=0$ block of the matrix contains (with reflection
symmetry included) about 242,000 states. However, there are other
symmetries that we did not use, such as point operations, that
commute with translations at the zone center and would further
reduce somewhat the dimension of the relevant space. {\em The
large value of not only the ground state, but also the lowest
excited state, overlap with the trial subspace is very strong
evidence that these systems are in the RR phase,} and would not
support an interpretation as hierarchy/composite-fermion states.

\begin{figure}
{\centering \includegraphics[width=3.0in]{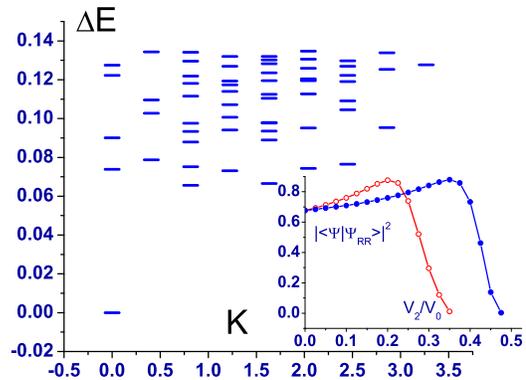} }
\caption{ \label{fig:specsphere} Low-lying spectrum for $21$
particles on the sphere for dipolar interaction with
$V_2/V_0=0.350$ vs.\ equivalent wavevector $K=L/R$ ($L$ is the
angular momentum). Inset shows the overlap-squared, with symbols
as in Fig.\ \ref{fig:overlaps}.}
\end{figure}

Fig.\ \ref{fig:specsphere} shows the spectrum in the spherical
geometry, with $V_2/V_0=0.350$. Again, a clear gap separates the
ground state from the rest of the spectrum. The results are
consistent with those on the torus.

\begin{figure}
{\centering \includegraphics[width=3.0in]{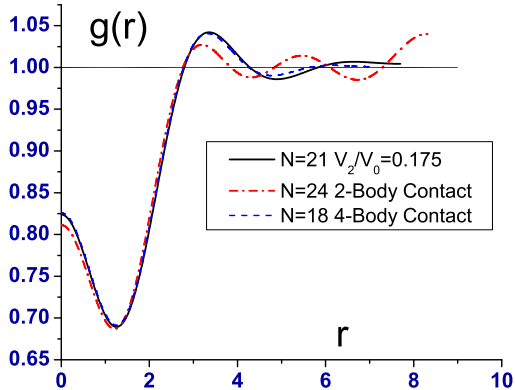} }
\caption{ \label{fig:pair} Two-particle correlation function
$g(r)$ of various ground states on the sphere, plotted against
great-circle distance $r$. $N=21$ $g(r)$ is for a $V_0$-$V_2$ model.}
\end{figure}

The two-particle correlation function $g(r)$ for the ground state
on the sphere is shown in Fig.\ \ref{fig:pair}, for the $V_0$,
$V_2$-only interaction for $21$ particles, together with that of
the RR trial state (the ground state of $H_4$) for $18$ particles,
for comparison. In general, $g(r)$ in a strongly-correlated state
should show a correlation hole at short distances (i.e.\
$g(0)<1$), and in an incompressible fluid state it should tend to
$1$ exponentially in $r$ at long distances. In the present case,
while $g(r)$ is less than 1 at $r=0$, it has a local maximum there
(mentioned in Ref.\ \cite{rj}, and this seems to be a general
feature of the ground states for $\nu>1/2$). This surprising
result may indicate a general tendency for bosons to cluster, at
least in these rotationally-invariant ground states. The behavior
at large $r$ suggests that these systems are on the verge of being
larger than the relevant correlation length. For the RR trial
state, $g(r)$ for $18$ particles has essentially converged to its
thermodynamic limit.

We can gain insight into the structure of $g(r)$ as $\nu$
increases by considering the vortex lattice. In this state there
are holes in the density at some fixed positions. Accordingly,
$g(r)$ in the vortex lattice, averaged over shifts and rotations
of the lattice to represent the finite-size system, will have
maxima and minima at arbitrarily large $r$; in particular, there
will be a maximum at $r=0$. It seems that in the RR series, the
oscillations in $g(r)$ die out with increasing $r$ at least for
small $k$, but that as $k$ increases the oscillations extend out
to larger $r$, and the maximum at $r=0$ becomes larger. We suspect
that for sufficiently large $k$, the RR trial states actually
exhibit a transition to vortex-lattice long-range order (or
possibly to one of the other ordered states found in Ref.\
\cite{crs}). This would be analogous to the Laughlin states at
$\nu=1/m$, which exhibit increasing oscillations with increasing
$m$, and eventually develop crystalline order of the particles
($g(0)=0$ in these cases) \cite{laugh}. Note that in both cases,
the long-range order in the trial states is (or would be)
occurring for the exact ground states of some family of special
Hamiltonians, that are not the ones of most physical interest, yet
the transition is of the same type as that for the latter (though
the critical $\nu$ may be much different, as for the Laughlin
states).

\begin{figure}
{\centering \includegraphics[width=3.0in]{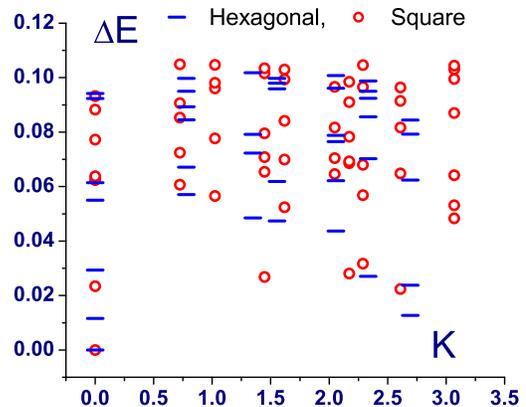} }
\caption{ \label{fig:deltaspecpbc} Low-lying spectrum (in units of $g$) for $18$
particles on the torus vs.\ pseudomomentum $K$, for $H_{\rm s}$
for two different unit cells.}
\end{figure}

Fig.\ \ref{fig:deltaspecpbc} shows the spectrum of the pure
contact interaction $H_{\rm s}$ (i.e.\ $V_2/V_0=0$) on the torus.
This, in conjunction with the corresponding $g(r)$ on the sphere
for $N=24$ particles shown in Fig.\ \ref{fig:pair}, shows no clear
signal that the system is incompressible. While it is possible
that the correlation length is simply larger, and that larger
systems would exhibit incompressibility, we cannot rule out the
possibility that the fluid is this region is compressible, or that
there is a breaking of translational symmetry in the thermodynamic
limit. Indeed, the dependence of the spectrum on the aspect ratio
in the torus geometry suggests the presence of a broken symmetry
``stripe'' phase competing with the RR phase \cite{crr}.

In conclusion, we have exhibited clear evidence that for bosons in
the lowest Landau level at filling factor $3/2$, when a moderate
amount of longer-range interaction is included, the ground state
is an incompressible fluid of a type that possesses non-Abelian
statistics for the quasiparticle excitations.

We wish to thank  David DeMille and Steven Girvin for discussions.
EHR thanks F.D.M. Haldane for use of his code in some of the calculations. 
We acknowledge support
from DOE contract no.\ DE-FG03-02ER-45981 (EHR), 
from NSF grant DMR0086191 at initial stages of the work (EHR), 
from NSF grant no.\ DMR-02-42949
(NR), and from EPSRC grant GR/S61263/01 (NRC).

\begin{references}
\bibitem{ellb} We use QH conventions, so that
$\ell_B=\ell_\perp/\sqrt{2}=\sqrt{\hbar/(2M\omega_\perp)}$, where
$\ell_\perp$ ($\ell_3$) is the quantum oscillator length for
motion in the $x$--$y$ plane (resp., $3$ direction) in the trap.
\end{references}
\end{document}